\documentstyle[prb,multicol,aps,epsf]{revtex}
\begin{document}
\title{Polariton Dispersion Law in Periodic Bragg and Near-Bragg Multiple Quantum Well Structures} \author{L.I. Deych}
\address{Department of Physics, Seton Hall University, 400 South Orange Ave, South\\
Orange, NJ 07079}
\author{A.A. Lisyansky}
\address{Department of Physics, Queens College of CUNY, Flushing, NY 11367}
\date{\today}
\maketitle

\begin{abstract}
The structure of polariton spectrum is analyzed for periodic multiple quantum well structures with periods  at or close to Bragg resonance condition at the wavelength of the exciton resonance. The results obtained used to discuss recent reflection and luminescent experiments by M. H\"{u}bner et al [Phys. Rev. Lett. {\bf 83}, 2841 (1999)] carried out with
long multiple quantum well structures. It is argued that the discussion of quantum well structures with large number of wells is more appropriate in terms of normal modes of infinite periodic structures rather then in terms of super- and sub- radiant modes. 
\end{abstract}

\begin{multicols}{2}
\section{Introduction}
Optical properties of excitons confined in quasi-two-dimensional quantum well (QW) structures attract a great deal of interest (see for review Ref. \onlinecite{Andreanireview}).  Starting from pioneering work by Agranovich and Dubovsky \cite{Agranovich}, it was understood that since the translational invariance in quasi-two-dimensional systems is broken in the direction normal to the plane of confinement, the coupling between excitons and light would lead to the radiative decay of excitons. This situation is usually described in terms of quasi-modes with complex eigen-energies. Imaginary parts of the latter characterize radiative  life-times of the respective modes.  Systems with multiple QW's (MQW's) demonstrate the presence of several quasi-modes with different radiative decay rates\cite{Andreani1995}.  For a few of those modes the radiative decay rates turn out to be larger than those for a single QW, and are actually growing with the number of QW's in the structure. Such modes are often called bright or super-radiant, while the modes with reduced radiative decay are called dark or sub-radiant\cite{Citrin1994,Bjork1995}. One of the theoretical and experimental methods to identify quasi-modes of MQW's is to consider reflection coefficient, which has complex-valued poles at the modes' frequencies\cite{Andreani1995}. The imaginary part of the frequency is interpreted as a half-width of the reflection resonance. 

The interpretation of optical properties of MQW's in terms of super- and sub-radiant modes gives a clear physical picture when the number of QW's is not very large. In systems with a larger number of wells, such an interpretation may be misleading. Consider, for instance, recent experiments described in Ref. \onlinecite{Khitrova1999}, where reflection and luminescence were studied for structures with up to 100 QW's. These experiments used the so called Bragg resonance structures, for which the period of the structure, $a$, satisfies the Bragg resonance condition, $a/2 =\lambda_{0}$, for the wavelength $\lambda_{0}$ of the radiation at the first heavy-hole exciton resonance frequency $\omega_0$. The theory of such structures in terms of super-radiant modes was developed in a number of papers\cite{Ivchenko1994,Ivchenko1998}. The main result of the theory is that there exists just one ``super-radiant" mode with a life-time $N\Gamma_0$, where $N$ is the number of the wells in the structure, and $\Gamma_0$ is a radiative life-time of excitons in a single well. The reflection coefficient from such a structure is given by\cite{Ivchenko1994}
\begin{equation}
R = \frac{(N\Gamma_0)^2}{\displaystyle{\left(\omega-\omega_0\right)^2+\left(\gamma +N\Gamma_0\right)^2}} ,
\label{reflection}
\end{equation}
where $\gamma$ is a homogeneous exciton broadening. This expression describes a very broad reflection resonance with the maximum at the Bragg resonance frequency. Eq. (\ref{reflection}) obviously brakes down when $N$ grows too large, but the interpretation of this equation in terms of the super-radiant mode becomes ambiguous even before that. In Ref. \onlinecite{Khitrova1999} the luminescence from a MQW structure with the number of wells up to $100$  was found to be very small at the frequency of the super-radiant mode. This seemingly paradoxical result becomes quite obvious if one considers the spectrum of MQW's in the superlattice limit. When the number of QW's increases, so called sub-radiant modes lose the imaginary component of their frequencies, and form regular stationary normal modes of an infinite periodic structure\cite{Ivchenko1998}. At the same time super-radiant modes become evanescent modes of the band-gaps of the structure. The reflection coefficient in band-gaps is close to one (if the homogeneous broadening is small enough), and its frequency dependence is very broad with almost rectangular shape. No propagating excitations exist at these frequencies, so it is obvious why the luminescence detected in Ref. \onlinecite{Khitrova1999} in this region was so weak. These rather straightforward discussion is warranted by the overuse of the terminology of super-radiance in the context of MQW's.

The experiments of Ref. \onlinecite{Khitrova1999} are the first where long MQW's with Bragg or near Bragg periods are studied. As just mentioned, it is more natural to discuss these experiments in terms of stationary excitations of an infinite periodic superlattice. Even though the dispersion equation for this system in its  general form has been obtained by many authors\cite{Citrin1994,Keldysh1988,Ivchenko1991,Andreani1994,Deutsch1995}, the detailed analysis of  this equation under Bragg or near Bragg conditions has not been carried out. To discuss details of the polariton dispersion in such a situation is the main objective of the present paper. The results of this discussion will be useful in better understanding the results of Ref. \onlinecite{Khitrova1999} and similar experiments. 

\section{The structure of the spectrum and polariton dispersion laws for a periodic Bragg superlattice}
 A general expression for the polariton dispersion law in a periodic QW superlattice was derived many times by different authors\cite{Citrin1994,Keldysh1988,Ivchenko1991,Andreani1994,Deutsch1995}. For a wave propagating in the direction of growth it has the following form:
\begin{equation}
\cos\left(Qa\right)=\cos\left(\frac{\omega}{c}a\right)-\frac{2\Gamma_0\omega}{\omega_0^2-\omega^2-2\gamma\omega}\sin\left(\frac{\omega}{c}a\right),
\label{DE}
\end{equation}
where $Q$ is the Bloch vector of the polariton and $c$ is the  speed of light in a background material. Generalization for an oblique direction is straightforward: $\omega/c$ is replaced with $k_z=\sqrt{(\omega/c)^2-k_{\|}^2}$, where $k_{\|}$ is an in-plane component of the wave vector. For short period superlattices, $a\omega/c\ll 1$, this equation is reduced to the standard polariton dispersion in a dispersionless material. In the absence of the homogeneous broadening, there is a polariton gap between $\omega_0$ and $\sqrt{\omega_0^2+ 4\Gamma_0 c/a}$. In general, band-gaps in the polariton spectrum are determined by inequalities:
\begin{eqnarray}
\frac{2\Gamma_0\omega}{\omega_0^2-\omega^2}\cot\left(\frac{\omega}{2c}a\right)&<&1, \label{0},\\
\frac{2\Gamma_0\omega}{\omega_0^2-\omega^2}\tan\left(\frac{\omega}{2c}a\right)&>&1, \label{pi}
\end{eqnarray}
where the polariton wave vector, $Q$ is 0 at the end of the interval determined by the first of this inequalities, and $Q=\pi/a$ at the ending point of the second one. 
For frequencies close to $\omega_0$ these inequalities are often solved approximately in the so called resonance approximation, where the frequency is taken equal to $\omega_0$ everywhere except for the exciton resonance denominator\cite{Ivchenko1998}. This approximation fails, however, for 
 Bragg structures satisfying the condition
\begin{equation}
\frac{a\omega_0}{c}=\pi
\label{Bragg}
\end{equation}
because the last term in Eq. (\ref{DE}) describing interaction between QW excitons and light vanishes at the exciton resonance frequency, $\omega_0$. In the absence of homogeneous broadening $\gamma$, the denominator in this term also vanishes, and, therefore, this case requires careful, albeit elementary, analysis. 

Inequalities (\ref{0}) and (\ref{pi}) in this case can be rewritten as 
\begin{eqnarray}
\frac{2\Gamma_0\omega}{\omega^2-\omega_0^2}\tan\left(\frac{\omega-\omega_0}{2c}a\right)&<&1,\label{0mod}\\
\frac{2\Gamma_0\omega}{\omega^2-\omega_0^2}\cot\left(\frac{\omega-\omega_0}{2c}a\right)&>&1.\label{pimod}
\end{eqnarray}
One can notice now that the first of these inequalities is never violated for frequencies close to $\omega_0$ as long as $\Gamma_0\ll\omega_0$. The boundaries of the band-gap are determined entirely by Eq. (\ref{pimod}), which means that at the both ends of the gap, the polariton wave vector $Q=\pi/a$. From Eq. (\ref{pimod}) we find that the polariton band-gap is determined by the inequalities
\begin{equation}
\omega_0-\sqrt{\frac{2\omega_0\Gamma_0}{\pi}}<\omega<\omega_0 + \sqrt{\frac{2\omega_0\Gamma_0}{\pi}},\label{gap}
\end{equation}
provided that inequality $\sqrt{\Gamma_0/\omega_0}\ll 1$ holds, which is usually true in real systems ($\sqrt{\Gamma_0/\omega_0}\sim 10^{-2}$ in the experiment of Ref.\cite{Khitrova1999}).

In the presence of homogeneous broadening the band-gap is not clearly defined, but it is remarkable that if $\gamma\not=0$ the solution of Eq. (\ref{DE}) at $\omega=\omega_0$ is real, $Q=\pi/a$, while, as Eq. (\ref{gap}) shows, this solution acquires an imaginary part when $\gamma=0$. In order to get a better understanding of the situation we have solved dispersion equation (\ref{DE}) in the presence of the homogeneous broadening for the frequencies satisfying Eq. (\ref{gap}). We found that the real part of the polariton's wave vector $Q'$ and its imaginary part $Q''$ have the following form:
\begin{equation}
|Q'-\pi|=Q'' = \frac{1}{a}\sqrt{\frac{\pi\Gamma_0\epsilon}{\gamma\omega_0}},  \label{reQ}
\end{equation}
for $ |\epsilon|\ll\gamma$, where $\epsilon=\omega-\omega_0$.
Eq. (\ref{reQ}) shows that for small $|\epsilon|$, the imaginary part of the polariton wave vector indeed becomes zero along with $\epsilon$, while farther away from the resonance frequency $\omega_0$, $|\epsilon|\gg \gamma$ 
$$
Q'' = \frac{1}{a}\sqrt{\frac{2\pi\Gamma_0}{\omega_0}},
$$
which is the expression one would obtain at $\omega=\omega_0$ in the absence of the homogeneous broadening. 

Eq. (\ref{reQ}) suggests a simple explanation of the results of luminescence experiments carried out in Ref. \onlinecite{Khitrova1999} with the exact Bragg structures. In this work, a  peak of the luminescence at the resonance frequency, $\omega_0$, right in the middle of the polariton gap was observed. One can relate this peak to the zeroing of the imaginary part of the polariton wave number $Q$. The width of the peak is determined by the homogeneous broadening parameter, $\gamma$. This observation can be used in order to validate the suggested explanation.
\section{Near-Bragg MQW structures}
One of the important experimental results of Ref. \onlinecite{Khitrova1999} is the observation of changes in the luminescence pattern with the change in the period of the MQW structure. In this section we examine how the spectrum of the MQW's evolves when it is tuned away from the exact Bragg resonance. We solve inequalities (\ref{0}) and (\ref{pi}) approximately for the frequency region $|\omega-\omega_B|a/c\ll 1$, where $\omega_B$ is the Bragg frequency defined as $\omega_B=\pi c/a$. In this approximation one finds that when the system is tuned away from the exact Bragg condition, $\omega_0=\omega_B$, the band-gap given by Eq. (\ref{gap}) divides into two gaps. If $\omega_0>\omega_B$ one has for the two gaps:
\begin{eqnarray}
\omega_2&<\omega&<\omega_B \label{gap1},\\
\omega_0-\frac{1}{2}\pi\Gamma_0\frac{\omega_0-\omega_B}{\omega_B}&<\omega<&\omega_1, \label{gap2}
\end{eqnarray}
where 
\begin{eqnarray}
\omega_1& =& \frac{\omega_0+\omega_B}{2}+\frac{1}{2}\sqrt{(\omega_0-\omega_B)^2+\displaystyle{\frac{16\Gamma_0\omega_B^2}{\pi(\omega_B+\omega_0)}}}, \nonumber \\
\omega_2& =& \frac{\omega_0+\omega_B}{2}-\frac{1}{2}\sqrt{(\omega_0-\omega_B)^2+\displaystyle{\frac{8\Gamma_0\omega_B}{\pi}}} . 
\end{eqnarray}
In the case of the detuning of the opposite sign, $\omega_0<\omega_B$, the band-gaps are determined by 
\begin{eqnarray}
\omega_2&<\omega&<\omega_0+\frac{1}{2}\pi\Gamma_0\frac{\omega_B-\omega_0}{\omega_B},\\
\omega_B &<\omega<&\omega_1 .
\end{eqnarray}
Using data from Ref. \onlinecite{Khitrova1999} ($\omega_0=1.491$ $eV$, $\Gamma_0=27$ $meV$), we can estimate positions of the gap boundaries for the system used in those experiments. The estimates are consistent with the positions of the luminescent peaks observed in Ref. \onlinecite{Khitrova1999} for different degrees of detuning. The general dispersion equation (\ref{DE}) can give the values of the wave numbers $Q$ corresponding to the modes excited in those experiments. We believe, however, that it is useful to have approximate ``long-wave" dispersion laws for those modes. For concreteness, we consider the case $\omega_0<\omega_B$. In this case, the excitations under interest belong to the branches with frequencies greater then  $\omega_0+\frac{1}{2}\pi\Gamma_0\frac{\omega_B-\omega_0}{\omega_B}\approx\omega_0$ and less than  $\omega_2$. The first of these branches approaches the band edge with $Q=0$, and the second one with $Q=\pi$. The near-the-edge dispersion laws for these branches can be obtained in the form:
\begin{equation}
\omega = \omega_0+\frac{1}{2}\pi\Gamma_0\frac{\omega_B-\omega_0}{\omega_B} + \pi\Gamma_0\frac{\omega_B-\omega_0}{8\omega_B}Q^2a^2,
\label{DEQ0}
\end{equation}
for the branch near $\omega_0$, and
\begin{equation}
\omega=\omega_2-\frac{(\omega_0-\omega_2)^3}{4\Gamma_0^2}\left(Qa-\pi\right)^2,
\label{DEQpi}
\end{equation}
for the branch near $\omega_2$.
One can see from these expressions that the effective masses of these two branches are significantly different. The one described by Eq. (\ref{DEQ0}) has a very small effective mass, and therefore the frequencies of this mode could only barely be distinguished from the resonance frequency $\omega_0$. The second branch, described by Eq. (\ref{DEQpi}), has much stronger dispersion, and, therefore, it must be separated from $\omega_0$ more strongly than by the width of the gap between $\omega_0$ and $\omega_2$. Indeed, using the numerical parameters of Ref. \onlinecite{Khitrova1999}, we find that the width of the gap for the detuning $\omega_0=0.98\omega_B$ is approximately equal to $1$ $meV$, while experimentally observed splitting between the modes is $3.2$ $meV$. This corresponds to the mode excited with a wave number $Q$ such that $|Qa-\pi|\approx 0.1$. The effective mass of this mode at the band edge under consideration, according to Eq. (\ref{DEQpi}), increases with an increase of detuning from the Bragg structure. This predictions can also be tested experimentally in order to check if the simple picture suggested in the present paper corresponds to the phenomenon observed in Ref. \onlinecite{Khitrova1999}.

Concluding, we analyzed the dispersion law of polaritons in periodic MQW structures at,or close to the Bragg resonance condition, $\omega_0a/c=\pi$, and established the pattern of band-gaps and conductivity bands arising in such structures. We also obtained analytical expressions for effective masses of polariton modes presumably observed in Ref. \onlinecite{Khitrova1999}. The theoretical results obtained were found to agree with experimental data. We also suggested some new experiments that can be used to further test the adequacy of the presented results. 

We wish to thank S. Schwarz for reading and commenting on the manuscript.  Work at Seton Hall University was supported by NATO Linkage Grant No 974573, work at Queens College was supported by PSC-CUNY research award.

\end{multicols}
\end{document}